\documentstyle[12pt]{article}
\begin{document}
\input epsf
\title{\bf Exclusive nonleptonic decays of $B$ mesons}
\renewcommand{\thefootnote}{\fnsymbol{footnote}}
\author{ D. Ebert\footnotemark[1]
~and R. N. Faustov\footnotemark[2]{}\ \footnotemark[3] \\
\small\it Institut f\"ur Physik, Humboldt--Universit\"at zu Berlin,\\
\small\it Invalidenstr.110, D-10115 Berlin, Germany\\
\\
V. O. Galkin\footnotemark[4] \\
\small\it Russian Academy of Sciences, Scientific Council for
Cybernetics,\\
\small\it Vavilov Street 40, Moscow 117333, Russia}

\date{}
\maketitle

\setcounter{footnote}{1}
\footnotetext{Supported in part by {\it Deutsche
Forschungsgemeinschaft} under contract Eb 139/1-2.}
\setcounter{footnote}{2}
\footnotetext{Supported by {\it Deutscher Akademischer
Austauschdienst} and
in part by {\it Russian Foundation for Fundamental Research}
 under Grant No.\ 96-02-17171.}
\setcounter{footnote}{3}
\footnotetext{On leave of absence from Russian Academy of Sciences,
Scientific Council for Cybernetics,
Vavilov Street 40, Moscow 117333, Russia.}
\setcounter{footnote}{4}
\footnotetext{Supported in part by {\it Russian Foundation for
Fundamental Research} under Grant No.\ 96-02-17171.}
\renewcommand{\thefootnote}{\arabic{footnote}}
\setcounter{footnote}{0}

\begin{abstract}
The energetic exclusive two-body nonleptonic decays of $B$ mesons are
investigated in the framework of the relativistic quark model within
the factorization approximation. The heavy quark expansion is used for
the calculation of form factors. The obtained results are in agrement
with available experimental data.

\medskip

\noindent PACS number(s): 13.25.Hw, 12.39.Ki
\end{abstract}

\section{INTRODUCTION}
The  investigation of exclusive nonleptonic decays of $B$ mesons
represents an important and complicated theoretical problem. In
contrast to the exclusive semileptonic decays, where  the weak current
matrix elements between meson states  are involved, nonleptonic decays
require the evaluation of  hadronic matrix elements of the local
four-quark operators. To simplify the analysis it is usually assumed
that the matrix element of the current-current weak interaction
factorizes into the product of two single current matrix elements.
Thus the problem reduces to the calculation of the meson form
factors, parametrizing the hadronic matrix elements of weak currents as
in the case of semileptonic decays, and the meson decay constants,
describing leptonic decays \cite{BSW}.
This makes the factorization hypothesis to be a very appealing
assumption. However, strong interaction effects, such as  final
state interactions, the rescattering  of the final hadrons etc.,
can  violate this approximation \cite{BSW,BBSUV}. There are also some
problems with the different renormalization point dependence of the
initial and factorized amplitudes \cite{DG,B}. Thus factorization can
not be considered as a universal approach to nonleptonic decays.

There were several theoretical developments which can help to justify
the factorization for certain nonleptonic decays of heavy mesons. It
has been shown in ref.~\cite{BGR} that factorization holds in the
limit of large number of colours $N_c$ in QCD. The leading $1/N_c$
corrections to this limit  have also been  considered. Moreover
intuitive arguments justifying factorization for the energetic
nonleptonic decays were given by Bjorken \cite{JB} on the basis of the
so-called colour transparency. In these decays the final hadrons are
produced in the form of point-like colour-singlet objects with a large
relative momentum. And thus the hadronization of the decay products
occurs  after they are too far away for strongly interacting with
each other, providing the possibility to avoid  final state
interactions. Dugan and Grinstein \cite{DG} discussed the
factorization hypothesis within the heavy quark effective theory. They
proved that factorization holds for the decays of a heavy $B$ meson
into a heavy $D$ meson and a light meson, where the light quarks,
which hadronize into a light meson, are highly energetic and
collinear. Therefore we have a good theoretical background to expect
that factorization can be applied to the consideration of
energetic nonleptonic decays of $B$ mesons.

In this paper we calculate the branching ratios of the exclusive
energetic nonleptonic decays of $B$ mesons in the framework of the
relativistic quark model on the basis of factorization. The
heavy-to-heavy hadronic form factors, appearing in the factorized
amplitudes, are constrained by the heavy quark effective theory (HQET)
\cite{N}. Our model explicitly satisfies all these constraints and
allows  the determination of the corrections in the inverse powers of
the heavy quark masses up to the second order \cite{8}. In the quoted
paper we have determined the Isgur-Wise function and the subleading
form factors in the hole kinematical range accessible in $B\to D$
transitions. We shall use these functions here to evolve $B\to D$
transition form factors from the point of zero recoil of the final $D$
meson to the  values of $q^2\approx m_f^2$, where $m_f$ is a mass of
the final light meson.  The form factors of the heavy-to-light
transitions have been calculated in our model at the point of maximum
recoil of the final light meson using the expansion in inverse powers
of the heavy $b$ quark mass from the initial $B$ meson and in inverse
powers of the large ($\sim m_b/2$) recoil momentum of the final light
meson \cite{semil}. We have also determined the $q^2$ dependence of
the form factors near this kinematical point. Thus we can calculate
the heavy-to-light form factors which are necessary for the
determination of energetic nonleptonic decay amplitudes. This
combination of the methods of heavy quark expansion and the
relativistic quark model increases the reliability of our predictions.
The comparison of the results with the available experimental data
will be the test of factorization.

The paper is organized as follows. In Sec.~2 we present the
expressions for nonleptonic decay amplitudes in the factorization
approximation. The relativistic quark model is described in Sec.~3. In
Sec.~4 the heavy-to-heavy transition form factors are discussed. The
heavy-to-light transition form factors are presented in Sec.~5.
Section~6 contain our results for the branching ratios of energetic
nonleptonic $B$ decays and their discussion. Our conclusions are given
in Sec.~7.

\section{NONLEPTONIC DECAY AMPLITUDES AND FACTORIZATION}

In the standard model $B$ decays are described by the effective
Hamiltonian, obtained by integrating out the heavy $W$-boson and top
quark and applying the operator product expansion.
For the case of $b\to c,u$ transitions,
\begin{eqnarray}
\label{heff}
H_{eff}&=&\frac{G_F}{\sqrt{2}}V_{cb}\left[c_1(\mu)O_1^{cb}+
c_2(\mu)O_2^{cb}\right] \cr
& & +\frac{G_F}{\sqrt{2}}V_{ub}\left[c_1(\mu)O_1^{ub}+
c_2(\mu)O_2^{ub}\right] +\dots,
\end{eqnarray}
where $V_{ij}$ are the corresponding Cabibbo-Kobayashi-Maskawa (CKM)
matrix elements. The Wilson coefficients $c_{1,2}(\mu)$ are evaluated
perturbatively at the $W$ scale and then they are evolved down to the
renormalization scale $\mu\approx m_b$ by the renormalization-group
equations. The ellipsis denotes the penguin operators, which Wilson
coefficients are numerically much smaller then $c_{1,2}$ \cite{SVZ}.
The local four-quark operators $O_1$ and $O_2$ are given by
\begin{eqnarray}
\label{o12}
O_1^{qb}&=& [({\tilde d}u)_{V-A}+({\tilde s}c)_{V-A}](\bar q
b)_{V-A}, \cr O_2^{qb}&=& (\bar qu)_{V-A}({\tilde d}b)_{V-A}+(\bar
qc)_{V-A}({\tilde s}b)_{V-A}, \qquad q=(u,c),
\end{eqnarray}
where the rotated antiquark fields are
\begin{equation} \label{ds}
\tilde d=V_{ud}\bar d+V_{us}\bar s,
 \qquad \tilde s=V_{cd}\bar d+V_{cs}\bar s,
\end{equation}
and for
the hadronic current the following notation is used
$$(\bar qq')_{V-A}=\bar q\gamma_\mu(1-\gamma_5)q' \equiv J_\mu.$$

The factorization approach to two-body nonleptonic decays $B\to
M_1(\bar q_1q'_1)M_2(\bar q_2q'_2)$ implies that the decay amplitude
can be approximated by the product of one-particle matrix elements:

\begin{eqnarray}
\label{factor} \langle M_1M_2|H_{eff}|B\rangle&=& \frac{G_F}{\sqrt{2}}
V_{q_1b}V_{q'_2q_2}\big[ a_1(\mu)\langle M_1|(\bar
q_1b)_{V-A}|B\rangle\langle M_2|(\bar q_2q'_2)_{V-A}|0\rangle \cr & &
+a_2(\mu)\langle M_2|(\bar q_2b)_{V-A}|B\rangle\langle M_1|(\bar
q_1q'_1)_{V-A}|0\rangle\big],
\end{eqnarray}
where
\begin{equation}
\label{amu} a_1(\mu)=c_1(\mu)+\frac{1}{N_c}c_2(\mu), \qquad
a_2(\mu)=c_2(\mu) +\frac{1}{N_c}c_1(\mu),
\end{equation}
$N_c$ is the number of colors ($N_c=3$).

In the general case, the renormalization point ($\mu$) dependence of
the product of current operator matrix elements does not cancel the
$\mu$ dependence of $a_i(\mu)$ or $c_i(\mu)$ \cite{DG,B}. Thus
non-factorizable contributions to (\ref{factor}) must be present in
order to make the physical amplitudes independent from the
renormalization scale $\mu$. However, as it is shown in \cite{DG}, in
the case of the production of an energetic light meson or meson
resonance it is posible to justify the factorization approximation and
the right-hand side of (\ref{factor}) is scale-independent. Thus we
limit our analysis of nonleptonic decays to consideration of
decays with at least one energetic meson in the final state (such as
$B\to D^{(*)}\pi(\rho)$ and $B\to \pi(\rho)\pi(\rho)$).

Before proceeding further, let us additionally note that in writing
eq.~(\ref{factor}) we discarded the contribution of the colour-octet
currents which emerge after the Fierz transformation of colour singlet
operators (\ref{o12}). Clearly these currents violate factorization
since they cannot provide transitions to the vacuum state. We also
neglected the so-called $W$-exchange and annihilation diagrams. In the
limit $M_W\to\infty$ they are connected by the Fierz transformation
and are doubly suppressed by the kinematic factor of order
$(m_D^2/m_B^2)$ and then dynamicly by the decreasing form factor
$F_{D\pi}(q^2=m_B^2)$ with $F_{D\pi}(0)=1$ (see ref.~\cite{BGR} for
details).

The coefficients (\ref{amu}) have been calculated at $\mu\approx m_b$
in the leading logarithmic approximation \cite{AM} as well as beyond
the leading logarithmic approximation \cite{B}. The result of
ref.\cite{B} is
\begin{equation}
\label{B} a_1=1.01\pm 0.02 \quad {\rm and}  \quad a_2=0.20\pm 0.05,
\end{equation}
which is close to the result of fitting experimental
data \cite{BH}
\begin{equation}
\label{aexp} a_1=1.03\pm 0.04\pm 0.06 \quad {\rm and} \quad
a_2=0.23\pm 0.01\pm 0.01.
\end{equation}
However the $a_2$ prediction (\ref{B}) is  renormalization scheme
dependent \cite{B}.

The matrix element of the current $J$ between the vacuum and final
pseudoscalar ($P$) or vector ($V$) meson states is parametrized by the
decay constants $f_{P,V}$
\begin{eqnarray}
\langle P|\bar q \gamma_\mu\gamma_5 q'|0\rangle&=&if_Pp_\mu, \cr
\langle V|\bar q\gamma_\mu q'|0\rangle&=&e_\mu m_Vf_V.
\end{eqnarray}

The matrix elements of the weak current $J$ between meson states have
the covariant decomposition \cite{BSW}:
\begin{eqnarray}
\label{ff1}
\langle P(p')|\bar q\gamma_\mu b|B(p)\rangle&=& \left[(p+p')_\mu-
\frac{m_B^2-m_P^2}{q^2}q_\mu\right]F_1(q^2)\cr
& & +\frac{m_B^2-m_P^2}{q^2}q_\mu F_0(q^2);
\end{eqnarray}
\begin{equation}
\label{ff2}
\langle V(p')|\bar q\gamma_\mu b|B(p)\rangle= \frac{2V(q^2)}{m_B+m_V}
i\epsilon_{\mu\nu\tau\sigma}e^{*\nu}p^\tau p'^\sigma;
\end{equation}
\begin{eqnarray}
\label{ff3}
\langle V(p')|\bar q\gamma_\mu\gamma_5 b|B(p)\rangle&=& (m_B+m_V)e^*_\mu A_1(q^2)-
\frac{A_2(q^2)}{m_B+m_V}(e^*q)(p+p')_\mu \cr
& & -2m_V\frac{(e^*q)}{q^2}q_\mu A_3(q^2) +2m_V\frac{(e^*q)}{q^2}q_\mu
A_0(q^2),
\end{eqnarray}
where $q=p-p'$ and $e$ is a polarization vector of the vector meson.
The form factor $A_3(q^2)$ is the linear combination
\begin{equation}
\label{ff4}
A_3(q^2)=\frac{m_B+m_V}{2m_V}A_1(q^2)-\frac{m_B-m_V}{2m_V}A_2(q^2),
\end{equation}
and in order to cancel the poles at $q^2=0$, it is necessary to
require
\begin{equation}
\label{ff5}
F_1(0)=F_0(0), \qquad A_3(0)=A_0(0).
\end{equation}
We calculate the corresponding form factors in the framework of the
relativistic quark model.

\section{RELATIVISTIC QUARK MODEL}

In the quasipotential approach a meson is described by the wave
function of the bound quark-antiquark state, which satisfies the
quasipotential equation \cite{3} of the Schr\"odinger type \cite{4}
\begin{equation}
\label{1}
{\left(\frac{b^2(M)}{2\mu_{R}}-\frac{{\bf
p}^2}{2\mu_{R}}\right)\Psi_{M}({\bf p})} =\int\frac{d^3 q}{(2\pi)^3}
 V({\bf p,q};M)\Psi_{M}({\bf q}),
\end{equation}
where the relativistic reduced mass is
\begin{equation}\mu_{R}=\frac{M^4-(m^2_a-m^2_b)^2}{4M^3};\end{equation}
and
\begin{equation}
{b^2(M) }
=\frac{[M^2-(m_a+m_b)^2][M^2-(m_a-m_b)^2]}{4M^2}.
\end{equation}
Here $m_{a,b}$ are quark masses; $ M$ is the meson mass and ${\bf p}$
is the relative momentum of quarks. While constructing the kernel of
this equation $V({\bf p,q};M)$ --- the quasipotential of
quark-antiquark interaction --- we have assumed that the effective
interaction is the sum of the one-gluon exchange term with the mixture
of long-range vector and scalar linear confining potentials. We have
also assumed that  the vector confining potential
contains the Pauli interaction. The quasipotential is defined by
\cite{5}:  \begin{eqnarray}V({\bf p,q},M)&=&\bar{u}_a(p)
\bar{u}_b(-p)\Big\{\frac{4}{3}\alpha_SD_{ \mu\nu}({\bf
k})\gamma_a^{\mu}\gamma_b^{\nu}\cr
& & +V^V_{\rm conf}({\bf k})\Gamma_a^{\mu}
\Gamma_{b;\mu}+V^S_{\rm conf}({\bf
k})\Big\}u_a(q)u_b(-q),
\end{eqnarray}
where $\alpha_S$ is the QCD coupling constant, $D_{\mu\nu}$ is the
gluon propagator; $\gamma_{\mu}$ and $u(p)$ are the Dirac matrices and
spinors and ${\bf k=p-q}$. The effective long-range vector vertex is
given by
\begin{equation}
\Gamma_{\mu}({\bf k})=\gamma_{\mu}+
\frac{i\kappa}{2m}\sigma_{\mu\nu}k^{\nu},
\end{equation}
where $\kappa$ is the Pauli interaction constant. Vector and
scalar confining potentials in the nonrelativistic limit reduce to
\begin{eqnarray}
V^V_{\rm conf}(r)&=&(1-\varepsilon)(Ar+B),\nonumber\\
V^S_{\rm conf}(r)& =&\varepsilon(Ar+B),
\end{eqnarray}
reproducing $V_{\rm nonrel}^{\rm conf}(r)=V^S_{\rm conf}+
V^V_{\rm conf}=Ar+B$, where
$\varepsilon$ is the mixing coefficient. The explicit expression for
the quasipotential with the account of  relativistic corrections of
order $v^2/c^2$ can be found in ref.\cite{5}.  All the parameters of
our model like quark masses, parameters of linear confining potential
$A$ and $B$, mixing coefficient $\varepsilon$ and anomalous
chromomagnetic quark moment $\kappa$ were fixed from the analysis of
meson masses \cite{5} and radiative decays \cite{6}. The quark masses
$m_b=4.88$ GeV; $m_c=1.55$ GeV; $m_s=0.50$ GeV; $m_{u,d}=0.33$ GeV and
parameters of the linear potential $A=0.18$ GeV$^2$ and $B=-0.30$ GeV
have standard values for quark models.  The value of the mixing
coefficient of vector and scalar confining potentials $\varepsilon=-1$
has been chosen from the consideration of the heavy quark expansion
\cite{8} and meson radiative decays \cite{6}, which are very sensitive
to the Lorentz-structure of the confining potential:  the resulting
leading relativistic corrections coming from vector and scalar
potentials have opposite signs for the radiative M1-decays \cite{6}.
Finally the universal Pauli interaction constant $\kappa=-1$ has been
fixed from the analysis of the fine splitting of heavy quarkonia ${
}^3P_J$- states \cite{5}.

The meson wave functions in the rest frame have been calculated by numerical
solution of the quasipotential equation (\ref{1}) \cite{14}. However,
it is more convenient to use analytical expressions for meson wave
functions. The examination of numerical results for the ground state
wave functions of mesons containing at least one light quark has shown
that they can be well approximated in the meson rest frame
by the Gaussian functions
\begin{equation}\label{Gauss}
\Psi_M({\bf p})
=\left({4\pi\over \beta_M^2} \right)^{3/4}\exp\left(-{{\bf
p}^2\over 2\beta_M^2}\right),
\end{equation}
with the deviation less than 5\%.

The parameters are

$$\beta_B=0.41\ {\rm GeV};\quad
\beta_D=0.38\ {\rm GeV};\quad \beta_{D_s}=0.44\ {\rm GeV};
\quad\beta_{\pi(\rho)}=0.31\ {\rm GeV}.$$

The matrix element of the local current $J$ between bound states in
the quasipotential method has the form \cite{7}
\begin{equation}
\label{8}
{ \langle  M' \vert J_\mu (0) \vert M\rangle
}= \int \frac{d^3p\, d^3q}{(2\pi )^6} \bar \Psi_{M'}^{b'}({\bf
p})\Gamma _\mu ({\bf p},{\bf q})\Psi_M^b({\bf q}),
\end{equation}
where $M(M')$ is the initial (final) meson, $\Gamma _\mu ({\bf p},{\bf
q})$ is the two-particle vertex function and  $\Psi_{M,M'}^{b,b'}$ are
the meson wave functions projected onto the positive energy states of
quarks and boosted to the moving reference frame.

This relation is valid for the general structure of the current
$J_{\mu}=\bar Q'G_{\mu}Q$, where $G_{\mu}$ can be an arbitrary
combination of Dirac matrices. The contributions to $\Gamma$ come from
Figs.~1(a) and 1(b).  Note that the contribution $\Gamma^{(2)}$ is the
consequence of the projection onto the positive-energy states. The
form of the relativistic corrections resulting from the vertex
function $\Gamma^{(2)}$ is explicitly dependent on the
Lorentz-structure of $q\bar q$-interaction.

The general structure of the current matrix element (\ref{8}) is
rather complicated, because it is necessary to integrate both with
respect to $d^3p$ and $d^3q$. The $\delta$-function in the expression
for the vertex function $\Gamma^{(1)}$ permits to perform
one of these integrations. As a result the contribution of
$\Gamma^{(1)}$ to the current matrix element has the usual structure
and can be calculated without any expansion, if the wave functions of
initial and final meson are known. The situation with the contribution
of $\Gamma^{(2)}$ is different. Here instead of the $\delta$-function
we have a complicated structure, containing the potential of the
$q\bar q$-interaction in a meson. Thus, in general case, we cannot
perform one of the integrations in the contribution of $\Gamma^{(2)}$
to the matrix element (\ref{8}). Therefore, it is necessary to use
some additional considerations. The main idea is to expand the vertex
function $\Gamma^{(2)}$ in such  a way that it
will be possible to use the quasipotential equation (\ref{1}) in order
to perform one of the integrations in the current matrix element
(\ref{8}). The realization of such expansion  differs for the cases of
heavy-to-heavy ($B\to D^{(*)}$) and heavy-to-light ($B\to \pi(\rho)$)
transitions.

\section{$B\to D^{(*)}$ DECAY FORM FACTORS}

In the case of the heavy-to-heavy ($B\to D^{(*)}$) meson decays we
have two natural expansion parameters, which are the heavy quark
masses ($m_b$ and $m_c$) in the initial and final meson. The most
convenient point for the expansion of vertex function $\Gamma^{(2)}$
in inverse powers of the heavy quark masses is
the point of zero recoil of the final $D$ meson, where ${\bf\Delta}=0$
(${\bf\Delta}={\bf p}_B-{\bf p}_{D^{(*)}}$).  It is easy to see that
$\Gamma^{(2)}$ contributes to the current matrix element at first
order of the $1/m_Q$ expansion. We limit our analysis to the
consideration of the terms up to the second order.  After the
expansion we perform the integrations in the contribution of
$\Gamma^{(2)}$ to the decay matrix element. As a result we get the
expression for the current matrix element, which contains the ordinary
mean values between meson wave functions and can be easily calculated
numerically. The results of such calculation are given in comparison
with the predictions of HQET \cite{N} in \cite{8}. Our model satisfies
all the constraints imposed on the form factors by heavy quark
symmetries and allows the determination of the Isgur-Wise and
subleading form factors \cite{8}.

The $q^2$ dependence of form factors at leading order of the $1/m_Q$
expansion is given by
\begin{eqnarray}
\label{hffq}
RF_1(q^2)=R\frac{F_0(q^2)}{1-\frac{q^2}{m_B+m_D}}=R^*V(q^2)& &\cr
= R^* A_0(q^2)
=R^*\frac{A_1(q^2)}{1-\frac{q^2}{m_B+m_{D^*}}}=R^*A_2(q^2)&=&\xi(w),
\end{eqnarray}
where
$$R^{(*)}=\frac{2\sqrt{m_Bm_{D^{(*)}}}}{m_B + m_{D^{(*)}}},$$
$$w=\frac{m_B^2+m_{D^{(*)}}^2-q^2}{2m_Bm_{D^{(*)}}}.$$
The Isgur-Wise function in our model is
\begin{equation} \label{iw} \xi(w)=\sqrt{\frac{2}{w+1}}
\exp \left(-\left(2\rho^2-\frac{1}{2}\right)\frac{w-1}{w+1}\right),
\end{equation}
with the slope parameter $\rho^2\simeq 1.02$, which is in accordance
with the recent CLEO II measurement \cite{BH}
$\rho^2=1.01\pm0.15\pm0.09$.

At the first order of the $1/m_Q$ expansion four additional
independent form factors arise in HQET \cite{N}. We have determined
these subleading form factors in the framework of our model \cite{8}
\begin{eqnarray}
\label{fo}
\xi_3(w) &=& (\bar\Lambda -m_q) \left(1+
\frac{2}{3}\frac{w-1}{w+1}\right)\xi(w), \nonumber\\
\chi_1(w)&=&\bar\Lambda\frac{w-1}{w+1}\xi(w), \nonumber\\
\chi_2(w)&=&-\frac{1}{32}\frac{\bar\Lambda}{w+1}\xi(w), \nonumber\\
\chi_3(w)&=&\frac{1}{16}\bar\Lambda\frac{w-1}{w+1}\xi(w),
\end{eqnarray}
where the HQET parameter $\bar \Lambda =M-m_Q$ in our model is equal
to the mean value of the light quark energy in the heavy meson $\bar
\Lambda=\langle \varepsilon_q\rangle\simeq0.54$ GeV.

We have also calculated the second order power corrections at the
point of zero recoil of the final meson \cite{8}. The obtained
structure of the second order corrections is in accord with
predictions of HQET \cite{FN}. As a result we got the values of the
$B\to D^{(*)}$ transition form factors at $q^2=q^2_{\rm max}$ up to
the second order terms. The higher order terms of the $1/m_Q$
expansion are negligibly small. However, for the consideration of the
energetic nonleptonic decays of $B$ mesons we need the form factor
values at $q^2=m_f^2\approx 0$ ($f=\pi,\rho\dots$ is a final light
particle).  Thus it is necessary to evolve the form factors from
$q^2_{\rm max}$ to $q^2\approx 0$. The corresponding $w$ range is not
very wide (from 1 to $\sim 1.6$). For such $w$ values the form factors
are dominated by the Isgur-Wise function \cite{GN}. Some small
contributions may arise from subleading form factors. The higher order
terms give very small corrections \cite{GN}.  Therefore we combine the
universal Isgur-Wise $q^2$ dependence of form factors (\ref{hffq})
with the subleading symmetry breaking corrections (\ref{fo}). The
resulting form factor $w$ dependence is shown in Figs.~2,3.

The values of form factors at $q^2=0$ are
\begin{eqnarray}
&&F_1(0)=F_0(0)=0.63, \quad V(0)=0.79,\cr
&&A_0(0)=0.63, \quad A_1(0)=0.62,\quad A_2(0)=0.61.
\end{eqnarray}

\section{$B\to\pi(\rho)$ DECAY FORM FACTORS}
In the case of heavy-to-light decays the final meson contains only
light quarks ($u$, $d$). Thus, in contrast to the heavy-to-heavy
transitions, we cannot expand matrix elements in inverse powers of the
final quark mass. The expansion of $\Gamma^{(2)}$ only in inverse
powers of the initial heavy quark mass at ${\bf\Delta}=0$ does not
solve the problem. However, the final  light meson has the large
recoil momentum, in comparison with its mass, almost in the whole
kinematical range. At the point of maximum recoil of the final light meson
the large value of recoil momentum $|{\bf\Delta}_{\rm max}|\sim m_b/2$
allows for the expansion of decay matrix element in $1/m_b$. The
contributions to this expansion come both from the inverse powers of
heavy $m_b$ from the initial $B$ meson and from inverse powers of the
recoil momentum $|{\bf\Delta}_{\rm max}|$ of the final light
$\pi(\rho)$  meson. In ref.\cite{semil} we carried out this expansion
up to the second order and performed one of the integrations in the
current matrix element (\ref{8}) using the quasipotential equation as
in the case of a heavy final meson. As a result we again get the
expression for the current matrix element, which contains only the
ordinary mean values between meson wave functions, but in this case at
the point of maximum recoil of the final light meson.

The found values of $B\to\pi(\rho)$ form factors at the point of
maximum recoil ($q^2=0$) are
\begin{eqnarray}
\label{lff}
&& F_1(0)=F_0(0)=0.21, \quad V(0)=0.29,  \cr
&& A_0(0)=0.18,   \quad A_1(0)=0.27, \quad A_2(0)= 0.30.
\end{eqnarray}

The $q^2$ behaviour of the heavy-to-light form factors near
$|{\bf\Delta}_{\rm max}|$ (corresponding to $q^2 =0$) is given by
\cite{semil}
\begin{eqnarray} \label{47}
F_1(q^2)&=&\frac{M_B+M_\pi}{2\sqrt{M_B
M_\pi}}\tilde\xi(w){\cal F}_+({\bf\Delta}_{\rm max}^2),\\
\label{46}
F_0(q^2)&=&\frac{2\sqrt{M_B M_\pi}}{M_B+M_\rho}\frac{1}{2}(1+w)
\tilde\xi(w){\cal F}_0 ({\bf\Delta}_{\rm max}^2),\\
\label{48}
A_1(q^2)&=&\frac{2\sqrt{M_B M_\rho}}{M_B+M_\rho}\frac{1}{2}(1+w)
\tilde\xi(w){\cal A}_1 ({\bf\Delta}_{\rm max}^2),\\
\label{49}
A_{0,2}(q^2)&=&\frac{M_B+M_\rho}{2\sqrt{M_B M_\rho}}\tilde\xi(w){\cal
A}_{0,2} ({\bf\Delta}_{\rm max}^2),\\
\label{50}
V(q^2)&=&\frac{M_B+M_\rho}{2\sqrt{M_B M_\rho}}\tilde\xi(w){\cal V}
({\bf\Delta}_{\rm max}^2),
\end{eqnarray}
where we have introduced the function \cite{semil}
\begin{equation}
\label{xi}
\tilde\xi(w)=\sqrt{2\over w+1}\exp
\left(-\eta\frac{\tilde\Lambda^2}{\beta^2_B}\frac{w-1}{w+1}\right),
\end{equation}
$$\eta =\frac{2\beta_B^2}{\beta_B^2+\beta_{\pi(\rho)}^2},
\qquad \tilde\Lambda\approx 0.53\ {\rm GeV}.$$
Equation (\ref{xi}) reduces to the Isgur-Wise function
(\ref{iw}) in the limit of infinitely heavy  quarks in the
initial and final mesons.

It is important to note that the form factors $A_1$ and $F_0$ in
(\ref{48}), (\ref{46}) have a different $q^2$ dependence than the
other form factors (\ref{47}), (\ref{49}), (\ref{50}). In  quark
models one usually assumes a pole \cite{16} or exponential
\cite{17} $q^2$ behaviour for all form factors. However, the recent
QCD sum rule analysis indicates that the form factor $A_1$ has a
$q^2$ dependence which is different from other form factors
\cite{18,19,ali}.

The extrapolation of the $q^2$ dependence (\ref{46})-(\ref{50}) to all
values of $q^2$ (or $w$), accessible in $B\to \pi(\rho)$ transitions,
introduces rather large uncertainties, because $w$ varies in a broad
kinematical range (from 1 to $\sim 19$ in $B\to\pi$ and from 1 to
$\sim 3.5$ in $B\to \rho$). However, the $w$ values for
$B\to\pi(\rho)$ form factors which  are really necessary for the
consideration of energetic nonleptonic decays are limited to a rather
small interval near $w_{\rm max}$ ($q^2=0$). Thus, the application of
formulas (\ref{46})-(\ref{50}) in this region is rather reliable. We
show the $w$ dependence of $B\to \pi(\rho)$ decay form factors in
Figs.~4,5.

\section{RESULTS AND DISCUSSION}

In the factorization approximation one can distinguish three classes
of $B$ meson nonleptonic decays (see Fig.~6) \cite{BSW}: the `class I'
transitions, such as $\bar B^0\to M_1^+ M_2^-$, where only the term
with $a_1$ in (\ref{factor}) contributes (i.e. both mesons are
produced by charged currents); `class II' transitions, such as $\bar
B^0\to M_1^0 M_2^0$, where only the term with $a_2$ in (\ref{factor})
contributes (i.e. both mesons are produced by neutral currents) and
`class III' transitions, such as $B^-\to M_1^0 M_2^-$, where both
terms can contribute coherently.

The results of the calculation of the nonleptonic branching ratios for
$B\to D^{(*)}$ and $B\to \pi(\rho)$ transitions are given in Tables~1
and 2 in comparison with other model predictions
\cite{NRSX}-\cite{Ch} and experimental data. The $B\to
D^{(*)}D_s^{(*)}$ decay branching ratios are presented for
completeness.

\begin{table*}[hbt]
\caption{Predicted branching ratios for  $B\to D^{(*)}M$ nonleptonic
decays in terms of $a_1$ and $a_2$
(in \%). Our model branching ratios are quoted for values of
$a_1=1.05$ and $a_2=0.25$ in comparison with experimental data (in
\%).  We use the values  $|V_{cb}|=0.038$ and $f_D=f_{D^*}=0.220$ GeV,
$f_{D_s}= f_{D_s^*}=0.260$ GeV, $f_{a_1}=0.205$ GeV [12] for our
estimates.}
\label{table:nonl1}
\small
\begin{tabular}{lcccc}
\hline
\hline
Decay & our result & \cite{NRSX}& our result& experiment \cite{BHP}\\
\hline
$\bar B^0\to D^+\pi^-$ & $0.29 a_1^2$ & $0.264 a_1^2$
& 0.32 & $0.31\pm0.04\pm0.02$\\
$\bar B^0\to D^+\rho^-$ & $0.79 a_1^2$ & $0.621 a_1^2$
& 0.87 & $0.84\pm0.16\pm0.05$\\
$\bar B^0\to D^{*+}\pi^-$ &$ 0.26 a_1^2$ & $0.254 a_1^2$
& 0.28 & $0.28\pm0.04\pm0.01$ \\
$\bar B^0\to D^{*+}\rho^-$ &$ 0.81 a_1^2$ & $0.702 a_1^2$
& 0.88 & $0.73\pm0.15\pm0.03$ \\
$\bar B^0\to D^+a_1^-$ & $0.78a_1^2$ & $0.673a_1^2$
& 0.86 & $0.60\pm0.22\pm0.24$  \\
$\bar B^0\to D^{*+}a_1^-$ & $0.92a_1^2$ & $0.970a_1^2$
& 1.02 & $1.27\pm0.30\pm0.05$\\
$\bar B^0\to D^{+}D_s$ & $1.37a_1^2$ &$1.213a_1^2$
& 1.51& $0.74\pm0.22\pm0.18$\\
$\bar B^0 \to D^+D_s^{*-}$ & $0.685 a_1^2$ & $0.859a_1^2$
& 0.75 & $1.14\pm0.42\pm0.28$ \\
$\bar B^0\to D^{*+}D_s^-$ & $0.82 a_1^2$ & $0.824a_1^2$
& 0.90 & $0.94\pm0.24\pm0.23$ \\
$\bar B^0\to D^{*+}D_s^{*-}$ & $2.50a_1^2$ & $2.203a_1^2$
& 2.75 & $2.00\pm0.54\pm0.49$ \\
$\bar B^0\to D^0\pi^0$ & $0.058a_2^2$ & $0.20a_2^2$
& 0.0036 & $<0.048$ \\
$\bar B^0 \to D^{*0}\pi^0$ & $0.056a_2^2$ & $0.21a_2^2$
& 0.0035 & $<0.097$ \\
$\bar B^0 \to D^0\rho^0$ & $0.053a_2^2$ & $0.14a_2^2$
& 0.0033 & $<0.055$ \\
$\bar B^0 \to D^{*0}\rho^0$ & $0.156a_2^2$ & $0.22a_2^2$
& 0.0098 & $<0.117$ \\
$B^-\to D^0\pi^-$ & $0.29(a_1+0.64a_2)^2$ & $0.265(a_1+1.230a_2)^2$
& 0.43 & $0.50\pm0.05\pm0.02$ \\
$B^-\to D^{*0}\pi^-$ & $0.27(a_1+0.69a_2)^2$ &$0.255(a_1+1.292a_2)^2$
& 0.40 & $0.52\pm0.08\pm0.02$ \\
$B^-\to D^0\rho^-$ & $ 0.81(a_1+0.36a_2)^2$ & $0.622(a_1+0.662a_2)^2$
& 1.06 & $1.37\pm0.18\pm0.05$ \\
$B^-\to D^{*0}\rho^-$ & $0.83(a_1^2+0.39a_2^2$ &
$0.703(a_1^2+0.635a_2^2$ & 1.17& $1.51\pm0.30\pm0.06$ \\
& $\phantom{0.83}+1.15a_1a_2)$ &$\phantom{0.703}+1.487a_1a_2)$ & &\\
$B^-\to D^0D_s^-$ & $1.40a_1^2$ & $1.215a_1^2$
& 1.55 & $1.36\pm0.28\pm0.33$ \\
$B^-\to D^0D_s^{*-}$ & $0.70a_1^2$ & $0.862a_1^2$
& 0.77 &$0.94\pm0.31\pm0.23$ \\
$B^-\to D^{*0}D_s^-$ & $0.84a_1^2$  & $0.828a_1^2$
& 0.92 & $1.18\pm0.36\pm0.29$ \\
$B^-\to D^{*0}D_s^{*-}$ & $2.56a_1^2$ & $2.206a_1^2$
& 2.80 & $2.70\pm0.81\pm0.66$ \\
\hline
\hline

\end{tabular}
\end{table*}

\begin{table*}[hbt]
\caption{Predicted branching ratios for $B\to\pi(\rho)M$ nonleptonic
decays. We use the experimental values for $f_\pi$, $f_\rho$
[12] and the value of $|V_{ub}|=0.0052$ [9] for our
model estimates. All numbers are branching ratios $\times 10^5$.}
\label{table:nonl2}
\begin{tabular}{lccccc}
\hline
\hline
Decay & our result & our result & \cite{Dean}&\cite{Ch}&
experiment UL \cite{PDG}\\
\hline
$\bar B^0\to\pi^+\pi^-$ & $0.331|V_{ub}|^2a_1^2$
& 0.99 & 1.8& & $<2.0$ \\
$\bar B^0\to\pi^+\rho^-$ & $0.857|V_{ub}|^2a_1^2$
& 2.55 & 4.8 &  & \\
$\bar B^0\to\rho^+\pi^-$ & $0.234|V_{ub}|^2a_1^2$
& 0.70 & 0.4 &  &  \\
$\bar B^0\to\pi^{\pm}\rho^{\mp}$ &$1.09|V_{ub}|^2a_1^2$
& 3.25& 5.2& & $<8.8$ \\
$\bar B^0\to\rho^+\rho^-$ & $ 0.794|V_{ub}|^2a_1^2$
& 2.36 & 1.3 & & $<220$\\
$\bar B^0\to\pi^0\pi^0$ & $ 0.17|V_{ub}|^2a_2^2$
& 0.028 & 0.06 & & $<0.91$\\
$\bar B^0\to\pi^0\rho^0$ & $ 0.54|V_{ub}|^2a_2^2$
& 0.092 & 0.14 &  &$<2.4$\\
$\bar B^0\to\rho^0\rho^0$ & $ 0.39|V_{ub}^2a_2^2$
& 0.067 & 0.05&  &$<28$\\
$\bar B^0\to D_s^-\pi^+$ & $0.285|V_{ub}|^2a_1^2$
& 0.85& 8.1 &1.9  &$<28$\\
$\bar B^0\to D_s^{*-}\pi^+$ & $1.06|V_{ub}|^2a_1^2$
& 3.1 & 6.1&2.7 & $<50$\\
$\bar B^0\to D_s^-\rho^+$ & $0.269|V_{ub}|^2a_1^2$
& 0.80 & 1.2 &1.0 & $<70$\\
$\bar B^0\to D_s^{*-}\rho^+$ & $2.25|V_{ub}|^2a_1^2$
& 6.7 & 4.5 &5.4 & $<80$\\
$B^-\to\pi^0\pi^-$ & $0.169|V_{ub}|^2(a_1+a_2)^2$
& 0.78 & 1.4 & & $<1.7$\\
$B^-\to\pi^0\rho^-$ & $0.438|V_{ub}|^2(a_1+0.52a_2)^2$
& 1.7 & 2.7 & & $<7.7$\\
$B^-\to\rho^0\pi^-$ & $0.120|V_{ub}|^2(a_1+1.95a_2)^2$
& 0.77 & 0.7& & $<4.3$\\
$B^-\to\rho^0\rho^-$ & $0.411|V_{ub}|^2(a_1+a_2)^2$
& 1.8 & 1.1& & $<100$\\
$B^-\to D_s^-\pi^0$ & $0.146|V_{ub}|^2a_1^2$
& 0.44& 3.9 &1.8  &$<20$\\
$B^-\to D_s^{*-}\pi^0$ & $0.545|V_{ub}|^2a_1^2$
& 1.6 & 3.0 &1.3 & $<33$\\
$B^-\to D_s^-\rho^0$ & $0.138|V_{ub}|^2a_1^2$
& 0.41 & 0.6&0.5 &$<40$\\
$B^-\to D_s^{*-}\rho^0$ & $1.16|V_{ub}|^2a_1^2$
& 3.4 & 2.4 &2.8 & $<50$\\
\hline
\hline
\end{tabular}
\end{table*}

We see that our results for the `class I' nonleptonic $B\to
D^{(*)}\pi(\rho)$ decays are close to the improved BSW model
predictions \cite{NRSX}, while our results for the `class II' and
$a_2$ contributions to `class III' decays are smaller than those of
\cite{NRSX}. These contributions come from $B\to\pi(\rho)$ transition
form factors, which have a different $q^2$ behaviour in our and the
BSW models. The BSW model assumes universal $q^2$ dependence of all
$B\to\pi(\rho)$ form factors. As already mentioned in our model we
find the $A_1$ and $F_0$ form factors to have a different $q^2$
dependence  than that of the other form factors (see
(\ref{46})-(\ref{50}) and Figs.~4,5). The form factor $F_0$ in our
model decreases with the growing of $q^2$ (decreasing of $w$) in the
kinematical range of interest for energetic nonleptonic decays (see
Fig.~4).  Note that our value for $B\to\pi$ form factors at $q^2=0$ is
approximately 1.5 times less than that of BSW, while the values for
$B\to\rho$ form factors are close in both models.

Our predictions for the branching ratios of $B\to D^{(*)}M$
nonleptonic decays presented in Table~1 agree with experimental data
within errors. Thus we can conclude that factorization works rather
well for `class I' and `class III' decays $B\to D^{(*)}\pi(\rho)$.
However, an improvement of experimental accuracy is needed to make a
definite conclusion. It will be very interesting to measure the `class
II' decay $\bar B^0\to D^{(*)0}\pi(\rho)^0$ branching ratios. Such
measurement will be the test of factorization for `class II'
nonleptonic decays and will help to constrain the $w$ (or $q^2$)
dependence of $B\to\pi(\rho)$ form factors.

We also present in Table~1 the predictions for $B\to D^{(*)}D_s^{(*)}$
nonleptonic decays, where only heavy mesons are present in the final
state. The factorization is less justified for such decays. However,
as we see from  Table~1  our predictions, based on the
factorization, are consistent with the experimental data for these
decays too.

For the branching ratios of $B\to\pi(\rho)M$ nonleptonic decays
presented in Table~2 there are only experimental upper limits at
present. The measurement of these decays will allow the determination
of the CKM matrix element $|V_{ub}|$, which is poorly studied. The
closest experimental upper limit to the theoretical predictions is for
the decay $\bar B^0\to\pi^+\pi^-$. It is approximately two times
larger than our prediction and is very close to the result of
\cite{Dean}. From this upper limit on  $B(\bar B^0\to\pi^+\pi^-)$
we get the limit on $|V_{ub}|$ in our model
$$ |V_{ub}|<7.4\times 10^{-3},$$
which is close to the value previously found from
semileptonic $B\to\pi(\rho)l\nu$ decays \cite{semil}:
$$|V_{ub}|=(5.2\pm1.3\pm0.5)\times 10^{-3}.$$

\section{CONCLUSIONS}
In this paper we have calculated the branching ratios of the energetic
exclusive nonleptonic decays of $B$ mesons on the basis of the
factorization approximation. In particular the form factors of $B\to
D^{(*)}$ and $B\to\pi(\rho)$ transitions have been evaluated using the
relativistic quark model and the heavy quark expansion. Such expansion
has been carried out up to the second order in the heavy quark masses.
Finally the momentum dependence of leading and subleading terms of
this expansion has been used for the determination of the
heavy-to-heavy transition form factors at $q^2=m_f^2$, where $f$ is a
final light meson.

The overall agreement of our predictions for two-body nonleptonic
decays of $B$ mesons with the existing experimental data
\cite{BHP,PDG} shows that the factorization approximation works
sufficiently well in the framework of our model. From another side, if
the factorization hypothesis is taken for granted, the  aforementioned
agreement confirms the selfconsistency of our approach, which
incorporates our previously obtained results for semileptonic and
leptonic decays of heavy mesons. In particular it would be quite
intersting to test the specific $q^2$ behaviour of the heavy-to-light
transition form factors $F_0$ and $A_1$ predicted by our model.
Another important problem is the determination of the coefficients
$a_1$ and $a_2$ via $c_1$ and $c_2$ directly from QCD. As it has been
discussed already in Sec.~2 their values found in ref.~\cite{B} are
close to those obtained from fitting experimental data, though one
should mention that $c_2$ is rather unstable with respect to the
renormalization scheme and scale. Nevertheless that means from our
point of view that the factorization hypothesis for energetic $B$
meson decays has more or less firm grounds within QCD, at least for
the `class I' decays.

The situation is essentially different for $D$ meson nonleptonic
decays. In this case the best fit to the experimental data yields
$a_1=1.26\pm0.10$ and $a_2=-0.51\pm0.10$ \cite{B,BH}. Meanwhile QCD
predicts \cite{B} (the instability of the coefficient here is even
stronger) $c_1=1.31\pm0.19$ and $c_2=-0.50\pm0.30$, which can only be
consistent with the result of fitting data if one drops the $1/N_c$
terms in eq.~(\ref{amu}) and puts $a_1\simeq c_1$ and $a_2\simeq c_2$.
Clearly, such an assumption would give a completely wrong result for
$B$ decays, namely a negative sign of $a_2$, which is ruled out by the
experimental data.  This result could indicate that the factorization
approach in $D$ decays is insufficient and non-factorizable
contributions there are large. In other words the $D$ meson is
possibly not heavy enough compared to the $B$ meson. The results of
refs.~\cite{DG} seem to point in the same direction.

The successful description of nonleptonic two-body decays of $B$
mesons makes the present approach appealing for the further
consideration of $B_{s,c}$ meson decays. Work in this direction is in
progress.

\medskip
\begin{center}
\Large \bf ACKNOWLEDGMENTS
\end{center}

We are grateful to M. Beyer, T. Feldman, M. Ivanov, J. K\"orner, T.
Mannel, S. Petrak, R. R\"uckl and B. Stech for useful discussions of
the results. One of us (R.~N.~F.) expresses his gratitude for the warm
hospitality extended to him at the Humboldt University of Berlin,
where the final part of this work was accomplished and DAAD for
financial support.

\newpage
\noindent {\Large \bf FIGURE CAPTIONS}

\bigskip

\noindent FIG.~1. {(a)~The lowest order vertex function
$\Gamma_\mu^{(1)}$.  (b)~The vertex function $\Gamma_\mu^{(2)}$ with
the account of the quark interaction. The dashed line corresponds to
the effective potential (17).  The bold line denotes the
negative-energy part of the quark propagator.}

\medskip

\noindent FIG.~2. The $w$ dependence of form factors of $B\to D$
transitions.

\medskip

\noindent FIG.~3.  The $w$ dependence of form factors of $B\to
D^*$ transitions.

\medskip

\noindent FIG.~4. The $w$ dependence of form factors of $B\to \pi$
transitions in the kinematical region of interest for the energetic
nonleptonic decays.

\medskip

\noindent FIG.~5. The $w$ dependence of form factors of $B\to \rho$
transitions in the kinematical region of interest for the energetic
nonleptonic decays.

\medskip

\noindent FIG.~6. Quark diagrams for two-body nonleptonic
$B$ decays: (a)~`class I' decays $\bar B^0\to M_1^+M_2^-$; (b)~`class
II' decays $\bar B^0\to M_1^0M_2^0$; (c)~`class III' decays $B^-\to
M_1^0M_2^-$.

\newpage
\begin{center}
\unitlength=0.65mm
\begin{picture}(150,150)
\put(10,100){\line(1,0){50}}
\put(10,120){\line(1,0){50}}
\put(35,120){\circle*{8}}
\multiput(32.5,130)(0,-10){2}{\begin{picture}(5,10)
\put(2.5,10){\oval(5,5)[r]}
\put(2.5,5){\oval(5,5)[l]}\end{picture}}
\put(0,118){$Q$}
\put(0,98){$\bar q$}
\put(0,107){$M$}
\put(65,118){$Q'$}
\put(65,98){$\bar q$}
\put(65,107){$M'$}
\put(43,140){$W$}
\put(31,85){\rm\bf (a)}
\put(10,20){\line(1,0){50}}
\put(10,40){\line(1,0){50}}
\put(25,40){\circle*{8}}
\put(25,40){\thicklines \line(1,0){20}}
\multiput(25,40.5)(0,-0.1){10}{\thicklines \line(1,0){20}}
\put(25,39.5){\thicklines \line(1,0){20}}
\put(45,40){\circle*{2}}
\put(45,20){\circle*{2}}
\multiput(45,40)(0,-4){5}{\line(0,-1){2}}
\multiput(22.5,50)(0,-10){2}{\begin{picture}(5,10)
\put(2.5,10){\oval(5,5)[r]}
\put(2.5,5){\oval(5,5)[l]}\end{picture}}
\put(0,38){$Q$}
\put(0,18){$\bar q$}
\put(0,27){$M$}
\put(65,38){$Q'$}
\put(65,18){$\bar q$}
\put(65,27){$M'$}
\put(33,60){$W$}
\put(90,20){\line(1,0){50}}
\put(90,40){\line(1,0){50}}
\put(125,40){\circle*{8}}
\put(105,40){\thicklines \line(1,0){20}}
\multiput(105,40.5)(0,-0.1){10}{\thicklines \line(1,0){20}}
\put(105,39,5){\thicklines \line(1,0){20}}
\put(105,40){\circle*{2}}
\put(105,20){\circle*{2}}
\multiput(105,40)(0,-4){5}{\line(0,-1){2}}
\multiput(122.5,50)(0,-10){2}{\begin{picture}(5,10)
\put(2.5,10){\oval(5,5)[r]}
\put(2.5,5){\oval(5,5)[l]}\end{picture}}
\put(80,38){$Q$}
\put(80,18){$\bar q$}
\put(80,27){$M$}
\put(145,38){$Q'$}
\put(145,18){$\bar q$}
\put(145,27){$M'$}
\put(133,60){$W$}
\put(70,5){\rm \bf (b)}

\end{picture}

{\bf FIG.~1}
\end{center}
\epsfysize=625pt  \epsfbox{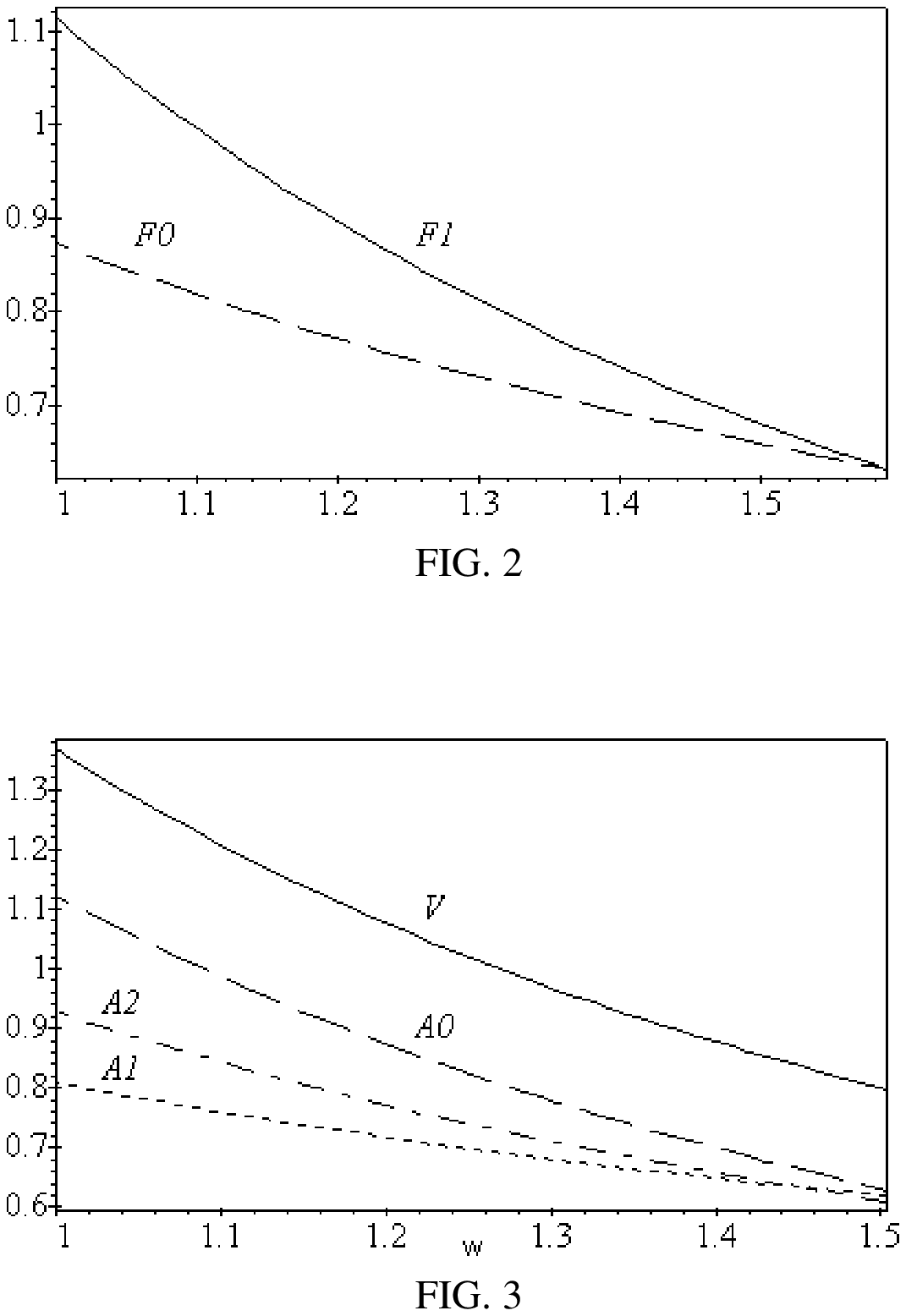}
\epsfysize=625pt  \epsfbox{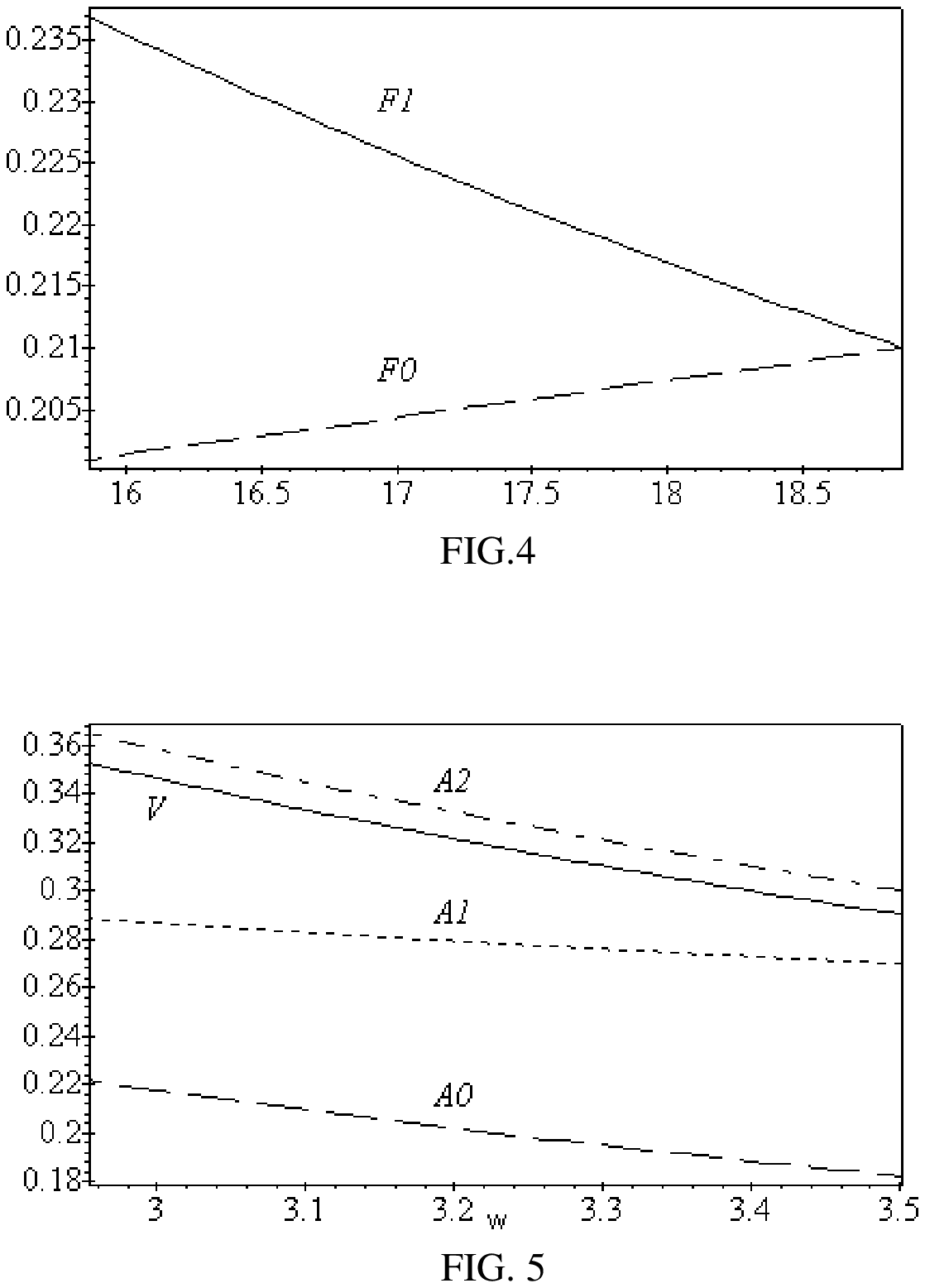}

\begin{figure}[htb]
\def\emline#1#2#3#4#5#6{%
       \put(#1,#2){\special{em:moveto}}%
       \put(#4,#5){\special{em:lineto}}}
\def\newpic#1{}
\unitlength=0.53mm
\large
\special{em:linewidth 0.4pt}
\linethickness{0.4pt}
\begin{center}
\begin{picture}(150.00,140.00)
\emline{40.00}{100.00}{1}{120.00}{100.00}{2}
\emline{40.00}{59.00}{3}{120.00}{59.00}{4}
\put(80.00,100.00){\circle{2.00}}
\emline{80.00}{100.00}{5}{78.00}{103.00}{6}
\emline{78.00}{103.00}{7}{84.00}{105.00}{8}
\emline{84.00}{105.00}{9}{81.00}{110.00}{10}
\emline{81.00}{110.00}{11}{87.00}{111.00}{12}
\emline{87.00}{111.00}{13}{84.00}{116.00}{14}
\emline{84.00}{116.00}{15}{88.00}{116.00}{16}
\put(88.00,116.00){\circle{2.00}}
\emline{120.00}{120.00}{17}{88.00}{116.00}{18}
\emline{88.00}{116.00}{19}{120.00}{140.00}{20}
\put(88.00,116.00){\circle{2.00}}
\put(80.00,100.00){\circle*{2.00}}
\put(88.00,116.00){\circle*{2.00}}
\put(70.00,110.00){\makebox(0,0)[cc]{$W$}}
\put(30.00,100.00){\makebox(0,0)[cc]{$b$}}
\put(30.00,59.00){\makebox(0,0)[cc]{$\bar d$}}
\put(130.00,100.00){\makebox(0,0)[cc]{$c,u$}}
\put(130.00,59.00){\makebox(0,0)[cc]{$\bar d$}}
\put(130.00,140.00){\makebox(0,0)[cc]{$\bar u,\bar c$}}
\put(130.00,120.00){\makebox(0,0)[cc]{$d,s$}}
\put(15.00,80.00){\makebox(0,0)[cc]{$\bar B^0$}}
\put(145.00,79.00){\makebox(0,0)[cc]{$M_1^+$}}
\put(145.00,130.00){\makebox(0,0)[cc]{$M_2^-$}}
\put(75.00,35.00){\makebox(0,0)[cc]{\bf(a) }}
\end{picture}

\begin{picture}(150.00,101.00)
\emline{40.00}{100.00}{1}{120.00}{100.00}{2}
\emline{40.00}{59.00}{3}{120.00}{59.00}{4}
\put(80.00,100.00){\circle*{2.00}}
\emline{80.00}{100.00}{5}{78.00}{97.00}{6}
\emline{78.00}{97.00}{7}{84.00}{96.00}{8}
\emline{84.00}{96.00}{9}{80.00}{92.00}{10}
\emline{80.00}{92.00}{11}{87.00}{91.00}{12}
\emline{87.00}{91.00}{13}{83.00}{87.00}{14}
\emline{83.00}{87.00}{15}{89.00}{86.00}{16}
\emline{89.00}{86.00}{17}{86.00}{82.00}{18}
\emline{86.00}{82.00}{19}{91.00}{81.00}{20}
\put(91.00,81.00){\circle*{2.00}}
\emline{120.00}{90.00}{21}{91.00}{81.00}{22}
\emline{91.00}{81.00}{23}{120.00}{70.00}{24}
\put(75.00,85.00){\makebox(0,0)[cc]{$W$}}
\put(30.00,100.00){\makebox(0,0)[cc]{$b$}}
\put(30.00,59.00){\makebox(0,0)[cc]{$\bar d$}}
\put(130.00,100.00){\makebox(0,0)[cc]{$c,u$}}
\put(130.00,89.00){\makebox(0,0)[cc]{$\bar u,\bar c$}}
\put(130.00,70.00){\makebox(0,0)[cc]{$d,s$}}
\put(130.00,59.00){\makebox(0,0)[cc]{$\bar d$}}
\put(150.00,95.00){\makebox(0,0)[cc]{$M_1^0$}}
\put(150.00,65.00){\makebox(0,0)[cc]{$M_2^0$}}
\put(15.00,80.00){\makebox(0,0)[cc]{$\bar B^0$}}
\put(75.00,35.00){\makebox(0,0)[cc]{\bf(b) }}
\end{picture}

\begin{picture}(300.00,140.00)
\emline{40.00}{100.00}{1}{120.00}{100.00}{2}
\emline{40.00}{59.00}{3}{120.00}{59.00}{4}
\put(80.00,100.00){\circle{2.00}}
\emline{80.00}{100.00}{5}{78.00}{103.00}{6}
\emline{78.00}{103.00}{7}{84.00}{105.00}{8}
\emline{84.00}{105.00}{9}{81.00}{110.00}{10}
\emline{81.00}{110.00}{11}{87.00}{111.00}{12}
\emline{87.00}{111.00}{13}{84.00}{116.00}{14}
\emline{84.00}{116.00}{15}{88.00}{116.00}{16}
\put(88.00,116.00){\circle{2.00}}
\emline{120.00}{120.00}{17}{88.00}{116.00}{18}
\emline{88.00}{116.00}{19}{120.00}{140.00}{20}
\put(88.00,116.00){\circle{2.00}}
\put(80.00,100.00){\circle*{2.00}}
\put(88.00,116.00){\circle*{2.00}}
\put(70.00,110.00){\makebox(0,0)[cc]{$W$}}
\put(30.00,100.00){\makebox(0,0)[cc]{$b$}}
\put(30.00,59.00){\makebox(0,0)[cc]{$\bar u$}}
\put(130.00,100.00){\makebox(0,0)[cc]{$c,u$}}
\put(130.00,59.00){\makebox(0,0)[cc]{$\bar u$}}
\put(130.00,140.00){\makebox(0,0)[cc]{$\bar u,\bar c$}}
\put(130.00,120.00){\makebox(0,0)[cc]{$d,s$}}
\put(15.00,80.00){\makebox(0,0)[cc]{$ B^-$}}
\put(145.00,79.00){\makebox(0,0)[cc]{$M_1^0$}}
\put(145.00,130.00){\makebox(0,0)[cc]{$M_2^-$}}

\emline{190.00}{100.00}{21}{270.00}{100.00}{22}
\emline{190.00}{59.00}{23}{270.00}{59.00}{24}
\put(230.00,100.00){\circle*{2.00}}
\emline{230.00}{100.00}{25}{228.00}{97.00}{26}
\emline{228.00}{97.00}{27}{234.00}{96.00}{28}
\emline{234.00}{96.00}{29}{230.00}{92.00}{30}
\emline{230.00}{92.00}{31}{237.00}{91.00}{32}
\emline{237.00}{91.00}{33}{233.00}{87.00}{34}
\emline{233.00}{87.00}{35}{239.00}{86.00}{36}
\emline{239.00}{86.00}{37}{236.00}{82.00}{38}
\emline{236.00}{82.00}{39}{241.00}{81.00}{40}
\put(241.00,81.00){\circle*{2.00}}
\emline{270.00}{90.00}{41}{241.00}{81.00}{42}
\emline{241.00}{81.00}{43}{270.00}{70.00}{44}
\put(225.00,85.00){\makebox(0,0)[cc]{$W$}}
\put(180.00,100.00){\makebox(0,0)[cc]{$b$}}
\put(180.00,59.00){\makebox(0,0)[cc]{$\bar u$}}
\put(280.00,100.00){\makebox(0,0)[cc]{$c,u$}}
\put(280.00,89.00){\makebox(0,0)[cc]{$\bar u,\bar c$}}
\put(280.00,70.00){\makebox(0,0)[cc]{$d,s$}}
\put(280.00,59.00){\makebox(0,0)[cc]{$\bar u$}}
\put(300.00,95.00){\makebox(0,0)[cc]{$M_1^0$}}
\put(300.00,65.00){\makebox(0,0)[cc]{$M_2^-$}}
\put(168.00,80.00){\makebox(0,0)[cc]{$ B^-$}}
\put(150.00,35.00){\makebox(0,0)[cc]{\bf(c) }}
\end{picture}
{\bf FIG. 6.}
\end{center}

\end{figure}


\begin{thebibliography}{99}
\bibitem{BSW} M. Bauer, B. Stech, and M. Wirbel, Z. Phys. C {\bf 34},
103 (1987).
\bibitem{BBSUV} I.I. Bigi {\it et al.}, in {\it B decays} edited by S.
Stone (World Scientific, Singapore, 1994) p.~132.
\bibitem{DG} M.J. Dugan and B. Grinstein, Phys. Lett. B {\bf 255}, 583
(1991), for more detailed discussion see C. Reader and N. Isgur,
Phys. Rev. D {\bf 47}, 1007 (1993).
\bibitem{B} A. Buras, Preprint MPI-Ph/94-60 (1994).
\bibitem{BGR} A.J. Buras, J.M. Gerard and R. R\"uckl, Nucl. Phys. B
{\bf 268}, 16 (1986).
\bibitem{JB} J.D. Bjorken, Nucl. Phys. B (Proc. Suppl.) {\bf 11}, 325
(1989).
\bibitem{N} M. Neubert, Phys. Rep. {\bf245}, 259 (1994).
\bibitem{8} R.N. Faustov and V.O. Galkin, Z. Phys. C {\bf 66}, 119
(1995).
\bibitem{semil} R.N. Faustov, V.O. Galkin and A.Yu. Mishurov, Phys.
Lett. B {\bf356}, 516 (1995), {\bf 367}, 391 (1996) (Erratum);
Phys. Rev. D {\bf 53}, 6302 (1996).
\bibitem{SVZ} M.A. Shifman, A.I. Vainshtein, and V.I. Zakharov, Nucl.
Phys. B {\bf 147}, 385, 448 (1979).
\bibitem{AM} G. Altarelli and L. Maiani, Phys. Lett. B {\bf 52}, 351
(1974); M.K. Gaillard and B.W. Lee, Phys. Rev. Lett. {\bf 33}, 108
(1974).
\bibitem{BH} T.R. Browder and K. Honscheid, Prog. Part. Nucl. Phys.
{\bf 35}, 81 (1995).
\bibitem{3} A.A. Logunov and A.N. Tavkhelidze, Nuovo Cimento {\bf29},
380 (1963).
\bibitem{4} A.P. Martynenko and R.N. Faustov, Teor.
Mat. Fiz. {\bf 64}, 179 (1985).
\bibitem{5} V.O. Galkin, A.Yu. Mishurov and R.N. Faustov, Yad. Fiz.
{\bf 55}, 2175 (1992).
\bibitem{6} V.O. Galkin and R.N. Faustov, Yad. Fiz. {\bf 44}, 1575
(1986); V.O. Galkin, A.Yu. Mishurov and R.N. Faustov, Yad.  Fiz.
{\bf 51}, 1101 (1990).
\bibitem{14} V.O. Galkin and R.N. Faustov, Teor. Mat. Fiz. {\bf85},
155 (1990).
\bibitem{7} R.N. Faustov, Ann. Phys. {\bf 78}, 176 (1973); Nuovo
Cimento {\bf 69}, 37 (1970).
\bibitem{FN} A.F. Falk and M. Neubert, Phys. Rev. D {\bf 47}, 2965
(1993).
\bibitem{GN} A.G. Grozin and M. Neubert, Preprint No. CERN-TH/96-144
(1996).
\bibitem{16} M. Wirbel, B. Stech and M. Bauer, Z. Phys. C {\bf29},
637 (1985).
\bibitem{17} N. Isgur, D. Scora, B. Grinstein and M.B.
Wise, Phys. Rev. D {\bf39}, 799 (1989).
\bibitem{18} P. Ball, Phys. Rev. D {\bf48}, 3190 (1993).
\bibitem{19} S. Narison, Report No. CERN-TH.7237/94 (1994).
\bibitem{ali} A. Ali, V.M. Braun and H. Simma, Z. Phys. C {\bf63},
437 (1994).
\bibitem{NRSX} M. Neubert, V. Rieckert, B. Stech and Q.P. Xu, in {\it
Heavy Flavours}, edited by A.J. Buras and H. Lindner (World
Scientific, Singapore, 1992), p. 286.
\bibitem{Dean} A. Deandrea {\it et al.}, Phys. Lett. B {\bf 318}, 549
(1993).
\bibitem{Ch} D. Choudury {\it et al.}, Phys. Rev. D {\bf 45}, 217
(1992).
\bibitem{BHP} T.E. Browder, K. Honscheid and D. Pedrini, Preprint
UH-515-848-96/OHSTPY-HEP-E-96-006 (1996), hep-ph/9606354~v2, to appear
in Ann. Rev. Nucl. \& Part. Sci. {\bf 46}.
\bibitem{PDG} R.M. Barnett {\it et al.}, Particle Data Group, Phys.
Rev. D {\bf 54}, 1 (1996).





\end{thebibliography}
\end{document}